\documentclass[10pt,conference]{IEEEtran}
\IEEEoverridecommandlockouts
\usepackage{cite}
\usepackage{booktabs}
\usepackage{amsmath,amssymb,amsfonts}
\usepackage{algorithmic}
\usepackage{graphicx}
\usepackage{textcomp}
\usepackage{xcolor}
\usepackage{framed}
\usepackage{float}
\usepackage{url}
\def\BibTeX{{\rm B\kern-.05em{\sc i\kern-.025em b}\kern-.08em
    T\kern-.1667em\lower.7ex\hbox{E}\kern-.125emX}}
\begin{document}

\title{What Happens When We Fuzz? Investigating OSS-Fuzz Bug History\\
}

\author{\IEEEauthorblockN{Brandon N. Keller}
\IEEEauthorblockA{\textit{Department of Software Engineering} \\
\textit{Rochester Institute of Technology}\\
Rochester, NY \\
bnk5096@rit.edu}
\and
\IEEEauthorblockN{Benjamin S. Meyers}
\IEEEauthorblockA{\textit{Department of Software Engineering} \\
\textit{Rochester Institute of Technology}\\
Rochester, NY \\
bsm9339@rit.edu}
\and
\IEEEauthorblockN{Andrew Meneely}
\IEEEauthorblockA{\textit{Department of Software Engineering} \\
\textit{Rochester Institute of Technology}\\
Rochester, NY \\
axmvse@rit.edu}
}

\maketitle

\newcommand{\todo}[1]{\textcolor{blue}{TODO: #1}} 
\newcommand{\readme}[1]{\textcolor{red}{README: #1}}

\newcommand{\rqonename}{Long-term Frequency}
\newcommand{\rqonetxt}{How does fuzzer-identified bug frequency change over time?}

\newcommand{\rqtwoname}{Response Time}
\newcommand{\rqtwotxt}{How long do developers take to respond once a fuzzer reports a bug?}

\newcommand{\rqthreename}{Lifespan}
\newcommand{\rqthreetxt}{How long did fuzzer-identified bugs exist within a codebase?}

\newcommand{\rqfourname}{Relevance}
\newcommand{\rqfourtxt}{Are fuzzers finding the same types of vulnerabilities that the project has historically had to fix?}

\newcommand{\rqfivename}{Learning Opportunities}
\newcommand{\rqfivetxt}{Do developers who introduce a fuzzer-identified bug end up fixing that same bug?}

\newcommand{\rqone}{RQ 1. \textbf{\rqonename} \rqonetxt }
\newcommand{\rqtwo}{RQ 2. \textbf{\rqtwoname} \rqtwotxt }
\newcommand{\rqthree}{RQ 3. \textbf{\rqthreename} \rqthreetxt}
\newcommand{\rqfour}{RQ 4. \textbf{\rqfourname}  \rqfourtxt}
\newcommand{\rqfive}{RQ 5. \textbf{\rqfivename} \rqfivetxt}

\newcommand{\methodstepzero}{\textbf{Step 0.} Investigate fuzzing practices of a single project}
\newcommand{\methodstepone}{\textbf{Step 1.} Collect fuzzbug data from the OSS-Fuzz issue tracker}
\newcommand{\methodsteptwo}{\textbf{Step 2.} Determine commit ranges from fuzzbug data}
\newcommand{\methodstepthree}{\textbf{Step 3.} Determine fix commits for each fuzzbug}
\newcommand{\methodstepfour}{\textbf{Step 4.} Determine origin commits for each fuzzbug}
\newcommand{\methodstepfive}{\textbf{Step 5.} Collect commit data on origin and fix commits}
\newcommand{\methodstepsix}{\textbf{Step 6.} Conduct analysis for research questions}

\begin{abstract}
BACKGROUND: Software engineers must be vigilant in preventing and correcting vulnerabilities and other critical bugs. In servicing this need, numerous tools and techniques have been developed to assist developers. Fuzzers, by autonomously generating inputs to test programs, promise to save time by detecting memory corruption, input handling, exception cases, and other issues.

AIMS: The goal of this work is to empower developers to prioritize their quality assurance by analyzing the history of bugs generated by OSS-Fuzz. Specifically, we examined what has happened when a project adopts fuzzing as a quality assurance practice by measuring bug lifespans, learning opportunities, and bug types.  

METHOD: We analyzed 44,102 reported issues made public by OSS-Fuzz prior to March 12, 2022. We traced the Git commit ranges reported by repeated fuzz testing to the source code repositories to identify how long fuzzing bugs remained in the system, who fixes these bugs, and what types of problems fuzzers historically have found. We identified the bug-contributing commits to estimate when the bug containing code was introduced, and measure the timeline from introduction to detection to fix. 

RESULTS: We found that bugs detected in OSS-Fuzz have a median lifespan of 324 days, but that bugs, once detected, only remain unaddressed for a median of 2 days. Further, we found that of the 8,099 issues for which a source committing author can be identified, less than half (45.9\%) of issues were fixed by the same author that introduced the bug.

CONCLUSIONS: The results show that fuzzing can be used to makes a positive impact on a project that takes advantage in terms of their ability to address bugs in a time frame conducive to fixing mistakes prior to a product release. However, the rate at which we find authors are not correcting their own errors suggests that not all developers are benefiting from the learning opportunities provided by fuzzing feedback. 
\end{abstract}

\begin{IEEEkeywords}
software security, fuzzing, vulnerability remediation, data mining
\end{IEEEkeywords}

\section{Introduction}
Developing software to be secure and reliable is a key responsibility of  software engineers. Vulnerabilities, for example, while important to find and fix, are a relatively small in size and in percentage of software bugs \cite{camilo_bugs_2015}. Modern software systems are complex, so any form of automation that helps developers find bugs faster might result in saving resources and lead to better software.

\emph{Fuzzing} is a software engineering practice that involves a tool (\textit{i.e.} a ``fuzzer'') automatically generating test data and executing tests with this generated data in search of exceptional conditions that lead to a crash. These conditions can include memory corruption, segmentation faults, and concurrency issues. Fuzzers have many promising benefits. They can run unobtrusively in the background on the system continuously as it is being developed, and any generated test data found to cause a crash can be re-run with every change to the system for regression testing. Fuzzers can potentially locate bugs faster than they might have ordinarily been found manually or after release. Further, the automated nature of fuzzers means that some bugs are found that a human may not be able to easily see, providing key learning opportunities for developers.

In recent years, the open source software development community has seen a widespread adoption of fuzzing as a practice. These fuzzers have been brought into prominent projects such as the linux init system systemd \cite{noauthor_systemd_2022}, openssh \cite{noauthor_openssh_2022}, curl \cite{noauthor_curl_2022}, and nginx \cite{noauthor_nginx_2022}. One fuzzing system, OSS-Fuzz \cite{oss-fuzz}, has been introduced to hundreds of open-source projects and provides robust integration with an issue tracking system, Monorail, so that developers can easily track their progress.

However, developers must balance their time and resources with all of the other demands from the software development lifecycle. Fuzzing does cost some time and effort in initial adoption, configuration, calibration of exceptional conditions, false positives, high priority fixes, and fixing low-severity issues. Development teams need to answer for themselves: is fuzzing worthwhile for our project? 

\textit{The goal of this work is to empower developers in determining if fuzzing is a worthwhile investment in their development process.} We investigated 44,102 issues that OSS-Fuzz published prior to March 12, 2022. We traced commits mentioned in the bug reports to their repositories and mined the repositories to measure their lifespan. 

We address the following research questions and discuss their specific relevance to software development teams.

\begin{itemize}
    \item[] \rqone
    \begin{itemize}
        \item[] Knowing how frequently a fuzzer will report issues allow for teams to estimate the impact on their development process. Furthermore, a decreasing frequency in the face of a growing project suggests that fuzzing is, in addition to detecting bugs, providing opportunity for developers to learn from their errors.
    \end{itemize}
    \item[] \rqtwo 
      \begin{itemize}
        \item[] Understanding the time it takes for developers to respond to bugs illustrates the ability for developers to make use of fuzzer feedback and to rectify concerns. A shorter time frame suggests that fuzzers provide feedback that is actionable enough to be addressed promptly.
    \end{itemize}
    \item[] \rqthree
      \begin{itemize}
        \item[] The duration for which a bug is present in a codebase provides evidence of the overall effectiveness of fuzzing. For example, short durations after adopting the practice suggest that the fuzzer is capable of locating bugs shortly after they are introduced.
    \end{itemize}
    \item[] \rqfour 
      \begin{itemize}
        \item[] Security vulnerabilities are among the most critical types of bugs that developers are responsible for preventing, finding, and fixing. Knowing what types of vulnerabilities have been found by fuzzers speaks to their overall relevance. In contrast, if fuzzers primarily find issues not relevant to the project, the practice might be a distraction that wastes resources. Fuzzers do more than security, but the secure software engineering community has provided a detailed and mature taxonomy in the Common Weakness Enumeration (CWE) to enable this type of analysis.
    \end{itemize}
    \item[] \rqfive
      \begin{itemize}
        \item[] When a developer is able to fix their own mistake, they are presented with an opportunity to learn from that mistake. A high rate of developers fixing mistakes that they introduced suggests that developers are being able to take feedback from their own work and to apply that to future development.
    \end{itemize}
\end{itemize}

\section{Background}
Throughout this paper, we use use the term \textbf{fuzzbug} to mean ``a bug automatically discovered by a fuzzer''. A fuzzbug can be a false positive, such as a miscalculated expected timeout or simulating an impossible situation. Fuzzbugs are often security-related, but not always. In relation to this paper, 21\% of fuzzbugs were deemed security-related by the OSS-Fuzz system. When a fuzzer finds a fuzzbug, it generates an issue on the Monorail tracker and saves the reproduction test case. This test case gets re-run periodically until it no longer produces an exceptional condition. This could be the result of developers intentionally fixing the fuzzbug, improving the system in a way that unintentionally fixes the fuzzbug, or removing the feature entirely.


\subsection{OSS-Fuzz}
OSS-Fuzz is a fuzzing system that combines many fuzzing techniques for use in open-source software \cite{oss-fuzz}. OSS-Fuzz is capable of detecting a wide variety of bugs and security concerns including memory leaks, memory overflows, and uncaught exceptions. As of April, 2022, developers of over 550 open-source projects have added OSS-Fuzz as part of their quality control mechanisms \cite{noauthor_oss-fuzz_2022}. In total, OSS-Fuzz has logged 46,043 issues to their issue tracking platform, Monorail.

Issues filed in Monorail are initially kept private. 30 days after an issue is corrected, the system makes the issue public. In cases where developers do not fix the issue within 90 days, the system also makes the issue public. Monorail entries detail the time at which the issue is detected with an associated commit range, the type of concern (\textit{i.e.} if it is Bug, Bug-Security, or a Build Failure), the crash state, the error type at the time of crash, the current status of the issue, and the point in time at which the issue was detected as fixed along with a commit range for this fix.

\subsection{Commit Ranges} \label{commit_ranges}
Commit ranges in Monorail (provided by OSS-Fuzz) point to a web page containing links to commit ranges in the corresponding project repository. For the purposes of this study, we only use the commit ranges belonging to a public GitHub repository. These GitHub commit ranges provide a collection of individual Git commits that were merged into the Git branch of analysis since the last time a test was performed and produced a different outcome.

\subsection{Vulnerabilities} \label{vulnerabilities}
In this work, we use data from the National Vulnerability Database (NVD)~\cite{noauthor_nvd_2022}, which is a comprehensive feed of vulnerabilities reported publicly from thousands of participating software projects. Each vulnerability in the NVD is given a Common Vulnerabilities and Exposures (CVE) number, and the reporter can apply some optional metadata as well. To convey a standard ``type'' of vulnerability, the CVE uses the Common Weakness Enumeration (CWE) to categorize vulnerabilities. The CWE is a taxonomy with a tree-like data structure containing over a thousand entries, including both broad categories and highly niche issues. 

\section{Related Work}
Our related work spans a variety of topics including empirical analysis of fuzzing, fuzzers more generally, and developer behavior as it pertains to vulnerability prevention and correction.

\subsection{Empirical Analysis of Fuzzing}
Previous work has studied fuzzbugs through an empirical lens. Ding \& Le Goues~\cite{ding_empirical_2021} performed an analysis of OSS-Fuzz data to understand the types of bugs that OSS-Fuzz reports, the time OSS-Fuzz takes to detect bugs, and the time it takes for a fix to be introduced to the project. Using the Monorail-provided data, they found that fuzzbugs had a median time to detect of 5 days and further that the median time for a bug to be fixed was 5.3 days. We perform similar analysis of OSS-Fuzz fuzzbugs, but supplement Monorail-provided data with analysis of project repositories and commits. This added source of information enables new research questions on  how developers respond to fuzzbugs and for identification of bug-contributing commits beyond the detection history of OSS-Fuzz.

Zhao \& Liu~\cite{zhao_empirical_2016} performed an empirical analysis of black-box mutational fuzzing, which is fuzzing for which inputs are derived from provided seed values. In this study, they investigated the diminishing returns of fuzzing in six projects and suggest mathematical models for predicting the results of fuzzing in terms of the bugs discovered over time. They found that in the projects explored, including popular projects like ffmpeg \cite{noauthor_ffmpeg_nodate} and autotrace \cite{noauthor_autotrace_nodate}, the amount of bugs discovered over time decreases. They concluded that these decreasing rates in discovery present a challenge to developers and expose a need for collaboration with ethical hackers to better combat bugs. Our work similarly investigated fuzzbug discovery rates, and looked for an increase or decrease in bug discovery rates, but in the context of a larger dataset including 562 projects.

\subsection{Fuzzers}
Developing fuzzers is a discipline in itself, and many techniques have been explored in depth across several studies. Man\'es \textit{et al.}~\cite{manes_art_2021} explored the design decisions that have gone into making modern fuzzers effective tools for bug detection through the survey of 11 years of fuzzing literature. Similarly, Wang \textit{et al.}~\cite{wang_systematic_2020} investigated developments in the fuzzing literature related to machine learning-based approaches. In another investigation of fuzzing literature, Fell~\cite{fell_review_2017} probed a variety of fuzzing tools and presented their strengths and weaknesses. In contrast with these surveys, we performed an empirical investigation of a single fuzzing system instead of evaluating a variety of fuzzing techniques.

Godefroid \textit{et al.}~\cite{godefroid_sage_2012} presented a case study of fuzzing at Microsoft with their SAGE white-box fuzzer. They found that SAGE, even when serving as the final step of the bug checking processes for Windows 7, found 1/3 of all discovered bugs. They concluded that SAGE has been invaluable for Microsoft and saved the company millions of dollars by preventing the need for additional security patches for the more than one billion Windows PCs. Similar to this work, our study investigates a single fuzzing system; however, we analyze OSS-Fuzz across 562 projects spanning dozens of organizations.

Metzman \textit{et al.}~\cite{metzman_fuzzbench_2021} detailed their fuzzing evaluation platform, FuzzBench. FuzzBench evaluates fuzzers by presenting them with a set of benchmark tests and evaluating their ability to detect bugs and to effectively provide coverage of a software system in these tests. Google, the developer of both OSS-Fuzz and FuzzBench, has used the results of these benchmark tests to select the fuzzers that are used in OSS-Fuzz. Further, Google has used fuzzbug data from OSS-Fuzz to develop benchmark tests for FuzzBench. Our work, instead of evaluating fuzzers based on coverage and ability to detect bugs, investigates fuzzers in terms of how they impact real-world projects and their ability to reduce fuzzbug lifespans. 

\subsection{Behavior in Vulnerability Correction \& Prevention}
Developers and their response to bugs have been explored in prior works. Munaiah \textit{et al.}~\cite{munaiah_natural_2017} investigated if the natural language used by developers during code reviews could explain why vulnerabilities are missed during these reviews. They found that reviews with low inquisitiveness and syntactic complexity, but higher sentiment were likely to miss bugs. In a similar work, Meyers \textit{et al.}~\cite{meyers_pragmatic_2019} explored security conversations related to bugs in the Chromium project. Their results indicate that the pragmatic characteristics (\textit{e.g.} politeness, formality, uncertainty) of developers' natural language play a role in security conversations and that there is potential for automated analysis to be used in determining effective communication strategies for discussing security concerns. While our work does not explore linguistics, the ways in which developers interact with bugs and their code is essential to understanding if fuzzing can be used as a teaching tool.

Outside of linguistic approaches, developer behavior has been studied through the measures of developer activity and code complexity. Shin \textit{et al.}~\cite{shin_evaluating_2011} investigated if software metrics, including those relating to churn, code complexity across numerous metrics, and developer activity, can be used to predict vulnerabilities. They found that these three classes of metrics, together, could predict vulnerable files that should be subject to further investigation. In another work, Islam \& Zibran~\cite{islam_what_2021} investigated 4,563 bug revisions across 5 projects. They found 38 edit-patterns used by developers in correcting bugs. Further, they found that if-blocks, especially those nested within loops or other if-blocks, are the most bug-prone structures across the projects they investigated. Our work builds upon these successes by exploring the ways that fuzzing can help developers to correct bugs and prevent future bugs.

\section{Methodology}
\label{sec:methodology}
Our methodology is intended to dig deep into the history of fuzzbugs, from the issue tracker to the repositories. All scripts used for this methodology are available for use (https://zenodo.org/badge/latestdoi/531572273).  Our methodology follows this process: 

\begin{enumerate}
    \item[] \methodstepzero
    \item[] \methodstepone
    \item[] \methodsteptwo
    \item[] \methodstepthree
    \item[] \methodstepfour
    \item[] \methodstepfive
    \item[] \methodstepsix
\end{enumerate}

\textbf{\methodstepzero}: This work began as a general investigation of the Linux systemd project in an effort to develop a security profile and to understand their practices for secure software engineering. We soon realized that among all of the quality and security control techniques/tools employed by developers of systemd, OSS-Fuzz presented a unique opportunity for a larger investigation of fuzzing as a practice. We closely examined post-release vulnerabilities of systemd in this process, as well as many of their individual fuzzbugs, to develop our research questions and data collection methodology. More details on this pilot study can be found in Section \ref{pilot}.

\textbf{\methodstepone}: 
Using a script we wrote with the Selenium library~\cite{noauthor_selenium_2022}, we collected all publicly-available OSS-Fuzz issues that are (1) not described as belonging to the ``Infra'' component, since these issues do not pertain to a specific project and are issues with OSS-Fuzz infrastructure instead, and (2) not tagged as a duplicate on the OSS-Fuzz Monorail issue tracker. Our script records the following details:
\begin{itemize}
    \item \textbf{Project:} The name of the corresponding project for each issue
    \item \textbf{Issue Number:} The unique identifier for an issue
    \item \textbf{Issue Status:} The current status of the issue (Open, Fixed, Won’t-Fix, \textit{etc.})
    \item \textbf{Detection Time:} The time when OSS-Fuzz detected the crash state
    \item \textbf{Fix Time:} The time when OSS-Fuzz repeated the error-causing test and found the problem to no longer be present
    \item \textbf{Type:} The type of issue detected--Bug, Build-Failure, or Bug-Security
    \item \textbf{Crash Type:} The specific error that caused the crash
    \item \textbf{Crash State:} The stack trace produced by the program when reaching the error or crash state
    \item \textbf{Fix Commit Range:} A link to commit ranges containing the fix (as determined by OSS-Fuzz)
    \item \textbf{Regression Commit Range:} A link to commit ranges where the original crash detection occurred (as determined by OSS-Fuzz)
\end{itemize}

We recorded all collected data in a comma-separated value (CSV) file for further processing.

\begin{figure} 
    \centering
    \includegraphics[width=200px]{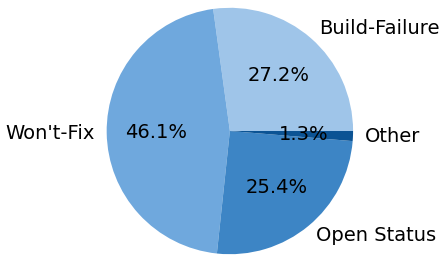}
    \caption{The breakdown of issues that did not have a Fix Commit Range detected when scraping Monorail issue pages}
    \label{fig:pie_breakdown}
\end{figure}

In total, we found and recorded 44,102 issues. Of the 44,102 issues we investigated, OSS-Fuzz labeled 13,155 of them as build-failures instead of a bug or otherwise did not provide fix commit details. We present the specific breakdown of the 13,155 excluded issues in \figurename~\ref{fig:pie_breakdown}.

\textbf{\methodsteptwo.} The fix commit range is the most important for the second stage of processing. As detailed in Section \ref{commit_ranges}, the commit ranges, in their default form, are not usable for analysis of individual commits. To retrieve the usable commit ranges from the provided web page, we wrote another Python script that uses Selenium \cite{noauthor_selenium_2022}.This script performs a check on each link provided on the web page to determine if any correspond to a GitHub repository. We add any commit ranges that correspond to a public GitHub repository to the CSV containing the data we acquired in the previous step of processing.

Of the 30,947 fix ranges that we investigated, we were able to map 26,216 to at least one GitHub commit range for use in the following stages of processing.

\textbf{\methodstepthree.} 
We used the GitHub API to inspect commit ranges to determine the most likely fix-contributing commit from within the range. We did this by inspecting both the functions and methods modified in the commit and the description of the commit. For each range, we inspected every commit, and sorted commits of relevance into categories. If we found a commit to contain a direct reference to the previously recorded OSS-Fuzz issue number, we considered it highly likely. Otherwise, we used the stack trace previously acquired to locate commits that modified the functions or methods recorded in the stack trace. We considered these commits to be the second most likely. If the commit range contained only a single commit, we considered this commit to be of the lowest likelihood. Once we evaluated all commits for a given issue, we labeled the commit of highest likelihood as the most likely contributor. If we found more than one commit of the same likelihood, we considered the first one we investigated of that likelihood to be the contributor as a default case. We recorded a link to the commit, the author of the commit, and the time of the commit in the CSV utilized in the previous steps. 

In total, we were able to map 10,129 of the 26,216 issues with a fix-commit range to a single fix commit. 

\textbf{\methodstepfour.} In the fourth stage of processing, we used the previously selected fix-contributing commit as the basis for an archeogit \cite{noauthor_archeogit_2021} powered \verb|git blame| operation. This tool, which utilizes the SZZ algorithm, enables the detection of the most likely bug contributing commit given a fix-commit hash. Multiple possible commits are provided by this tool. We selected the commit deemed most likely by archeogit to be the bug-contributing commit utilized for further operations. We performed this operation on all issues for which a fix-contributing commit was identified.

In total, we successfully located a bug-contributing commit for 8,827 of the 10,129 issues that had a valid fix commit. 

\textbf{\methodstepfive.} In the fifth stage of processing, we again utilized the GitHub API with a Python script to collect additional details of the bug-contributing commit as determined in Step 4. Specifically, we recorded the commit link, the author, and the time of the commit in the same CSV from prior stages of processing. 

We successfully retrieved the required information from the GitHub API for 8,114 of the 8,827 issues that we previously mapped to a bug-contributing commit.

\textbf{\methodstepsix.}
The methodology regarding the analysis of the data collected in the previous steps is detailed in Section \ref{Analysis}. Our data is available for download and use (https://zenodo.org/badge/latestdoi/531572273). 

\section{Data Analysis} \label{Analysis}

\subsection{Pilot Investigation: systemd} \label{pilot}

While our end-goal was to conduct a large-scale empirical study, we began our investigation with a detailed study of a single, representative project. Our intent with this pilot study is to gain a deeper understanding of the idiosyncrasies of the data, which then informed our next steps. We used this opportunity to find as much historical evidence of quality assurance practices as we could find. This open-ended stage was critical to crafting our research questions, honing our data collection methodology, and refining our analysis plan.

We conducted our pilot study by examining the security practices of systemd. We chose systemd because it is a prominent open source project with critical implications on Linux distributions worldwide. The project has a typical and healthy history of tracking and diligently fixing vulnerabilities. The systemd artifacts provided enough traceability between bugs, vulnerabilities, fixes, and documentation for an initial investigation.

The systemd project involves 1,627 Git committers, with the most contributions from Lennart Poettering who has contributed over 16,000 commits since the project began in 2010. Even on commits he is not authoring, Poettering is present on many of the pull requests and conversations in the project. 

The systemd project uses a variety of quality control practices designed to find vulnerabilities early. Since 2013, they employ static analysis by Coverity run on a daily basis and track the results. They make extensive use of unit testing, as noted in their code quality guide~\cite{noauthor_code_2022} and the pervasive existence of unit tests in the code base. Integration tests also run regularly, although with less documentation and guidelines than unit tests, to test systemd on various operating system distributions via nspawn~\cite{nspawn} and qemu~\cite{qemu}. They employ continuous integration and use Address Sanitizer~\cite{address_sanitizer} and Undefined Behavior Sanitizer~\cite{behave_sanitizer} on all incoming pull requests. Fuzzing is also one of their practices. The developers actively respond to OSS-Fuzz fuzzbugs filed on Monorail. 

We began the initial investigation of systemd's OSS-Fuzz data by collecting the systemd issues filed to Monorail (139 issues when we collected in September, 2021). OSS-Fuzz data shows that systemd adopted fuzzing as a regular practice in 2018. From visual inspection of the data, several trends were immediately apparent. Most notably, over 100 of the fix-contributing commits were made within a week of the relevant bug being detected. Also of note was that only a small subset of contributors regularly made fix-contributing commits. Additionally, our visual inspection of the data suggested that the fuzzer detected fewer bugs over time with only 21 bugs discovered in 2020, but 45 in 2019 and 58 in 2018. These realizations led to further questions about if these trends could be generalized across other projects that use OSS-Fuzz.

\subsection{\rqonename}
\textit{RQ1: \rqonetxt}

We can examine the value in utilizing a fuzzer by identifying the overall trends in fuzzbug discovery. Over time, developers are able to use the feedback from fixing fuzzbugs to learn from their mistakes and reduce their rate of bug creation. If fuzzing is intended to improve the quality of a software project, does that mean that fuzzbug discoveries decrease over time?

To measure this, we used the Detection Time, collected in Step 1, to determine the number of issues OSS-Fuzz filed each day over the entire lifetime of each project. This is similar to the methodology used in Zhao \& Liu~\cite{zhao_empirical_2016}. We then calculated a linear regression trendline for each project such that a positive slope would indicate an increasing frequency in fuzzbugs and that a negative slope would indicate a decreasing frequency of fuzzbugs. All 562 projects in the dataset had multiple issues that allowed us to use a regression analysis. 

All of the projects we studied are active and growing projects. Therefore, an increase in fuzzbugs may simply mean a increase in the size of the project, and would require further investigation into how fuzzbug rates compare when normalized for project size. However, a \emph{decreasing} fuzzbug frequency, despite the project growing in size, would suggest that the fuzzbug discovery rate is decreasing as well. 

Additionally, when a project adopts fuzzing as a practice, the fuzzer will find bugs that pre-existed in system. Thus, one would expect an initial batch of fuzzbugs to be found, followed by a stream of fuzzbugs later on. To account for this, we repeated the above analysis but removed bugs that had an origin commit that pre-dated the Project Adoption Date. We define Project Adoption Date as the Detection Date of the first fuzzbug filed for that project.

\subsection{\rqtwoname}
\textit{RQ2: \rqtwotxt}

The time that developers take to respond to reported fuzzbugs provides insight into how fuzzers impact a project. A fuzzbug discovery can be a sudden redirection of the development team's resources, which could be costly and result in slower responses to other concerns. A faster response time suggests that developers take fuzzbugs seriously and prioritize their correction, whereas slow response times may indicate that a difficult fix is required, that the developers do not seriously address fuzzbugs, or that the fuzzer has found a bug outside of the scope of the project.

To measure response time, we located the fix-contributing commit for each fuzzbug and used the time of this commit and the Detection Time, defined in Step 1, acquired directly from Monorail. We calculated response time, in days, by subtracting the detection time from the fix-contributing commit time. We excluded from consideration, for the purposes of these calculations, 68 issues because they related to a timeout error and had a fix originating in the OSS-Fuzz repository, indicating that the bug was not in the project itself. After excluding these 68 issues, 10,061 issues were left in the dataset for consideration of this research question.

\subsection{\rqthreename}
\textit{RQ3: \rqthreetxt}

The lifespan of a bug indicates, to large extent, how many versions of an application have been put at risk. This is especially important in the case of bugs that make software vulnerable. Li \& Paxson~\cite{li_large-scale_2017} found in their large scale 2017 study that, for bug lifespans, the median of project medians was 438 days for bugs that posed a security risk and even higher for bugs unrelated to security. Therefore, a median lifespan under 438 days would suggest that fuzzers allow for bugs to be located faster than otherwise, and a median lifespan greater than or close to 438 days would suggest that fuzzers make little or no impact in terms of bug discovery speed. 

To measure this, we used the fix-contributing and bug-contributing commits acquired in Steps 3 and 4 respectively and the data regarding these commits acquired in Step 5. 


We calculated lifespan in days by subtracting the time of the bug-contributing commit from the time of the fix-contributing commit. We excluded 68 issues from consideration due to their fix-contributing commits belonging to the OSS-Fuzz project and the Crash Type, as defined in Step 1, being a timeout as these characteristics suggest an issue with the fuzzer itself rather than the project of analysis. After removing these issues, 8,048 issues were left in the dataset for consideration of this research question.

\subsection{\rqfourname} \label{analysis-4}
\textit{RQ4: \rqfourtxt} 

Fuzzing promises to help in discovering exceptional conditions that may lead to security concerns within a software system. Therefore, there should be overlap between the bugs found by OSS-Fuzz and those reported in the form of CVEs. A high percentage of overlap between these two bug sources suggests that OSS-Fuzz is capable of being a valuable asset as it relates to ensuring software security. A low percentage overlap presents many possibilities including OSS-Fuzz being targeted towards a wider variety of bugs than merely those that have security-related concerns. Knowing the overlap between these two datasets can also help development teams calibrate their expectations when determining which quality assurance practices to adopt.

To measure this, we first pulled the National Vulnerability Database (NVD) data feed~\cite{noauthor_nvd_2022}. (See Section \ref{vulnerabilities} for background on vulnerability data.) We filtered the list of CVEs to include only those regarding the projects that utilize OSS-Fuzz. Optionally provided by CVE reporters, the CWE indicates the classification of the bug that contributed to the CVE. We mapped OSS-Fuzz Crash Types as described in Step 1 to any relevant CWEs as reported in the same project's set of CVEs.

We mapped 31 of 80 unique OSS-Fuzz Crash Types to at least one CWE. For some entries, such as null-dereference, OSS-Fuzz does not consider it a security risk where as CWE does. In those cases, we were inclusive and used the CWE classification. Thus, we effectively ignored the OSS-Fuzz classification of \texttt{Bug-Security} vs \texttt{Bug} and used the mapped CWEs for this stage of analysis. Our mappings are shown in Table~\ref{tab:mappings}. For ease of reading, we exclude any Crash Types for which there was no match and represent Crash Types for which there are many variations as a single generalized entry in this table.

\begin{table}[h]
\centering
\caption{The map of OSS-Fuzz crash types to relevant CWEs as also cited by relevant CVEs. Crash types without mappings have been left out for brevity.}
\label{tab:mappings}
\begin{tabular}{l l}
\toprule
\textbf{OSS-Fuzz Crash Type} & \textbf{CWE(s)} \\
\midrule
ASSERT                                                        & 670,617        \\ 
Bad-cast                                                      & 704,681,843     \\ 
Container-overflow READ                                       & 119,125         \\ 
Container-overflow WRITE                                      & 119,787         \\ 
Divide-by-zero                                                & 369             \\ 
Dynamic-stack-buffer-overflow READ                            & 119,125,788,121 \\ 
Dynamic-stack-buffer-overflow WRITE                           & 119,787,121     \\
Float-cast-overflow                                           & 681,1901 704    \\ 
Global-buffer-overflow READ                                   & 119,125,788,121 \\
Global-buffer-overflow WRITE                                  & 119,787,121     \\ 
Heap-buffer-overflow READ                                     & 119,125,788,121 \\ 
Heap-buffer-overflow WRITE                                    & 119,787,121     \\ 
Heap-double-free                                              & 415             \\ 
Heap-use-after-free READ                                      & 416             \\
Heap-use-after-free WRITE                                     & 416             \\ 
Index out of range                                            & 129             \\
Index-out-of-bounds                                           & 129             \\
Integer divide by zero                                        & 369             \\ 
Integer-overflow                                              & 190             \\ 
Null-dereference                                              & 476             \\ 
Null-dereference READ                                         & 476             \\ 
Null-dereference WRITE                                        & 476             \\ 
Potential-null-reference                                      & 476             \\ 
Slice bounds out of range                                     & 129             \\ 
Stack-buffer-overflow WRITE                                   & 119,787,121     \\ 
Stack-buffer-underflow READ                                   & 119,125,788,121 \\ 
Stack-buffer-underflow WRITE                                  & 119,787,121     \\ 
Uncaught exception                                            & 248             \\ 
Unsigned-integer-overflow                                     & 190             \\ 
Use-of-uninitialized-value                                    & 908,1188       \\ 
\bottomrule
\end{tabular}
\end{table}

To compare CWE entries of CVEs and fuzzbugs, we must account for the complex, multi-typed hierarchical nature of the CWE taxonomy. A single CWE entry can be, from high-level to low-level, a ``category'', ``pillar'', ``class'', ``base``, or a ``variant``. Bases, Variants, and Classes can all be children of each other, with multiple inheritance, where Pillar is the highest form of abstraction in the CWE. Category is a cross-cutting way of grouping CWE entries and is generally not useful for identifying a vulnerability. Thus, having different CWE identifiers does not necessarily mean that they are not the ``same type'' of problem. One person may apply a category and another may apply a child of that category, for instance. The CWE guidelines provide help in navigating this \cite{CWEGuidelines}, but CVE reporters are not guaranteed to read the guidelines. This issue is a challenge to many researchers looking to apply CWE \cite{Alqahtani2017, Islam2017} to existing vulnerabilities.

We accounted for this challenge in several ways. Typically, one would follow the guidelines of ``lowest-level mapping should be performed when possible''\cite{CWEGuidelines}. In this study, erring on the side of inclusion, we mapped each fuzzbug type to a CWE identifier. We then looked up the ancestry of that identifier, stopping at the Pillar level. That way, if a CVE reporter used a broader ancestor of type Class, Base, or Variant, we considered that a match. This choice helps to avoid a ``splitting hairs'' problem. For example, the subtleties of the Base entries of \textit{CWE-266 Incorrect Privilege Assignment} and \textit{CWE-271 Privilege Dropping / Lowering Errors} might not be consistently differentiated between by CVE reporters. Thus, when a CVE and a fuzzbug are the ``same type'', we define that as having the CWE of the CVE be in the ancestry of the fuzzbug CWE.

We also note that, just because a fuzzer can find, for example, \textit{CWE-787 Out-of-bounds Write} in a project does not mean that the specific CVEs of that type \textit{could} have been found. We are only comparing types to err on the side of inclusion.

\subsection{\rqfivename}
\textit{RQ5: \rqfivetxt}

Fuzzers, rather than automatically correcting or mitigating an issue, merely identify a problem for developers to act upon. This approach creates an opportunity for a developer to learn from and correct their own mistakes. Alternatively, development teams can suffer from a ``throw it over the wall'' \cite{se-adages2017} syndrome where there is a bifurcation between those who create new features and those who maintain those features. A large percent of issues being addressed by the original author suggests that developers are taking advantage of the opportunities OSS-Fuzz presents. A small percent may indicate that developers are not given the opportunity to learn in this style or that they are instead given opportunities to fix and learn from issues created by others. Analysis of the pilot study presented in Section \ref{pilot} suggests that there may instead be a select group of developers tasked with correcting issues.

To measure the rate at which developers are able to correct their own mistakes, we first located the bug-contributing and fix-contributing commits as described in Steps 4 and 3 respectively. We then use the authorship information acquired in Step 5 and compare each of the two commits by the author's email. We consider a bug to be fixed by the person that introduced the bug if these emails match. We excluded any issues with a fix-commit located in the OSS-Fuzz project and a Crash Type, as defined in Step 1, of timeout as this indicates the issue is with OSS-Fuzz and not the project of analysis. After this removal process, we processed 8,033 issues that were successfully associated with an author's email address.

\section{Results}
This section covers the results of the analysis described in Sections \ref{sec:methodology} and \ref{Analysis}.
\subsection{\rqonename}

\textit{RQ1: \rqonetxt}

\begin{figure*}[t]
    \centering
    \includegraphics[width=\textwidth]{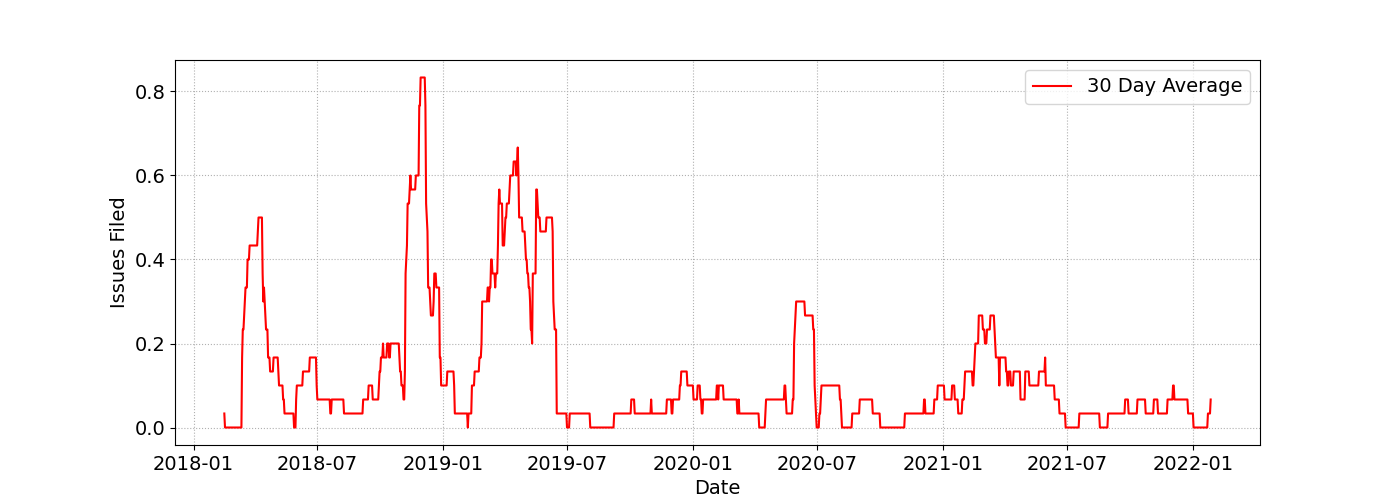}
    \caption{Linux systemd fuzzbug issues filed over time, 30-day rolling average}
    \label{fig:systemd}
\end{figure*}

Overwhelmingly, we found a downward trend in fuzzbug discoveries over time. Of the 562 total projects, 532 (94.6\%) of projects indicated a decreasing frequency of issues. As an example, we show a 30-day rolling average for issues filed for the systemd project in \figurename~\ref{fig:systemd}.

After adjusting to exclude those issues introduced prior to the start of fuzzing operations, of the 8,114 issues that could be mapped to an origin commit in Step 4, 3,813 (47\%) of these issues originated prior to the introduction of OSS-Fuzz to the project. With this new filtered dataset, we found that only 536 projects had enough corresponding issues to perform a linear regression, and that of these 536, only 278 (51.9\%) showed a decreasing bug frequency.

\begin{framed}
\noindent 94.6\% of the projects reported a downward trend in fuzzbug discovery over time, 51.9\% after adjusting for pre-existing fuzzbugs.    
\end{framed}

These results indicate that adopting a fuzzer can lead to an initial batch of pre-existing fuzzbugs. Additionally, an ongoing maintenance, often less than the initial batch, effort should be expected as more pre-existing and newly introduced fuzzbugs are discovered as development continues.

\subsection{\rqtwoname}
\textit{RQ2: \rqtwotxt}

We found some response time calculations resulted in negative values. This indicates that the fuzzbug was corrected prior to OSS-Fuzz's discovery of it. This means that the fix was manually detected and committed to a different branch, and not merged into the branch under observation by OSS-Fuzz when the fuzzer detected the bug on that branch.

We used the median of response times to examine the overall response time, as it is not as sensitive to outliers as mean. To adjust for outlier projects, we also compute a ``median of medians'', which is the median of every project's median of response times.

\begin{figure} 
    \centering
    \includegraphics[width=\columnwidth]{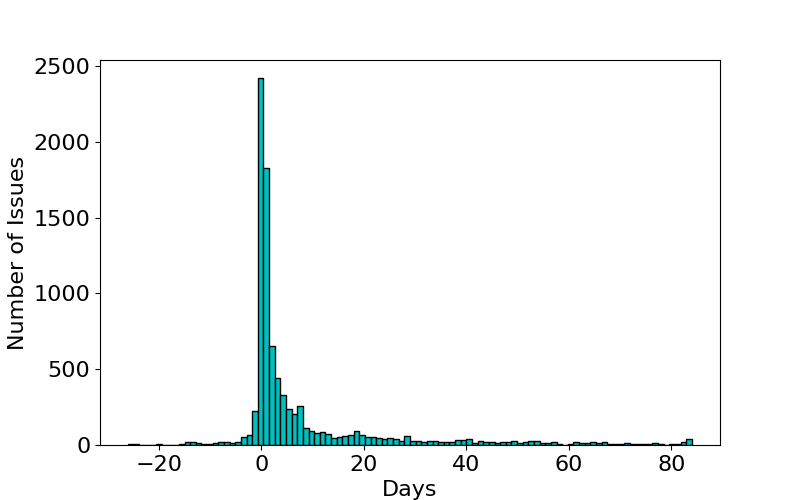}
    \caption{Histogram of 10,061 response times, fuzzer detection until fix commit, omitting 1,200 issues to increase visibility of the quickest response times.}
    \label{fig:response_hist}
\end{figure}

We found that, overall, the median time for a developer to respond to a fuzzbug was 2 days. When adjusting to use a median of project medians, we found that the median increases to only 2.5 days. This time to respond is less than half of what was found by Ding \& Le Goues~\cite{ding_empirical_2021} as the commit history revealed additional context that suggests developers are quick to respond to bugs, but there may be delays in getting a fix merged into the project or otherwise put into the branch of analysis. A histogram showing the distribution of response times across all issues is found in \figurename~\ref{fig:response_hist}. 

\begin{framed}
\noindent The median time for a developer to respond to a fuzzbug is 2 days. 74.5\% of fuzzbugs are responded to within 7 days.
\end{framed}

These results indicate that, with response times being relatively fast, the feedback the fuzzer is providing is actionable and serious enough to be dealt with.

\subsection{\rqthreename}
\textit{RQ3: \rqthreetxt}
\begin{figure} 
    \centering
    \includegraphics[width=\columnwidth]{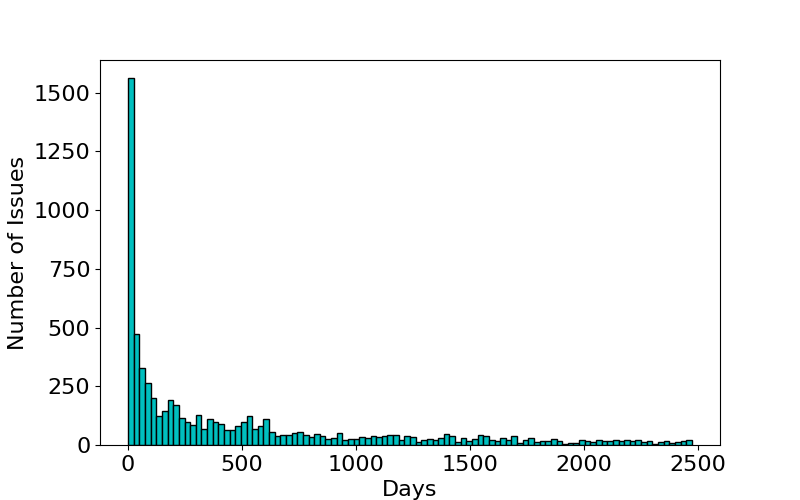}
    \caption{Histogram of fuzzbug lifespan from the time of source commit creation until the time of fix commit creation. 1,200 issues are omitted for clarity.}
    \label{fig:lifespan_hist}
\end{figure}

We found an overall median lifespan of fuzzbugs of 324 days. When considering a median of project medians, we found the median fuzzbug lifespan to be 303.5 days. Both of these medians, including both those fuzzbugs related to security and those not related to security, are more than 100 days fewer than the 438 days for security related bugs found by Li \& Paxson~\cite{li_large-scale_2017}. This suggests that OSS-Fuzz has helped to reduce bug lifespans. We show a histogram of calculated lifespans in \figurename~\ref{fig:lifespan_hist}.

\begin{framed}
\noindent The median lifespan of a fuzzbug is 324 days, 303.5 days when considering a median of project medians. 59\% of issues have a lifespan less than or equal to 365 days.
\end{framed}

Vulnerability lifespan is shorter for fuzzbugs than for the bugs studied in \cite{li_large-scale_2017}, so fuzzers do find problems faster than in other projects. However, these results also indicate that even bugs detected by fuzzers can stay in the system for a relatively long time before being discovered, meaning that a lack of fuzzer feedback initially is not necessarily a guarantee of quality.

\subsection{\rqfourname}
\textit{RQ4: \rqfourtxt}
Table \ref{tab:catspills} shows the CWE entries encountered in CVE data in this portion of the study.  Table \ref{tab:catspills} includes the pillars and categories that were too broad to map to any OSS-Fuzz Crash Type. Table \ref{tab:cweclasses} shows the CWE identifiers, grouped at the Class level for visual clarity, for this study. The column ``Fuzzbug?'' refers to whether or not one of those identifiers is in Table \ref{tab:mappings}.

We found that 18,794 fuzzbugs belonged to a project that also had at least one CVE with a provided CWE. Of these 18,794 issues, only 1,731 (9.2\%) fuzzbugs matched a CVE from their project. 

\begin{framed}
\noindent 9.2\% of OSS-Fuzz issues are similar to a post-release vulnerability from the same project
\end{framed}

These results indicate that fuzzers are far from being comprehensive in the types of vulnerabilities they find and should not be solely relied upon as the main source of quality assurance against vulnerabilities.

Additionally, while one may consider 9.2\% to be a small overlap, a single vulnerability of those mentioned in Table~\ref{tab:mappings} can have highly severe consequences. The lack of overlap does not mean that fuzzing is not worthwhile, just more niche than one might expect.

\begin{table}[h]
\centering
\caption{CWE Categories and Pillars encountered in vulnerability type comparison }
\label{tab:catspills}
\begin{tabular}{l l}
\toprule
\textbf{CWE ID} & \textbf{CWE Name} \\
\midrule
\hline
\multicolumn{2}{c}{\textbf{CATEGORIES}} \\
\hline
CWE-19 & Data Processing Errors   \\
CWE-16 & Configuration    \\
CWE-388 & Errors    \\
CWE-255 & Credentials Management Errors   \\
CWE-189 & Numeric Errors    \\
CWE-264 & Permissions, Privileges, and Access Controls    \\
CWE-320 & Key Management Errors   \\
CWE-254 & Security Features   \\
CWE-275 & Permission Issues   \\
CWE-417 & Communication Channel Errors    \\
CWE-399 & Resource Management Errors    \\
CWE-310 & Cryptographic Issues    \\
CWE-17 & Code (DEPRECATED)    \\
CWE-18 & Source Code (DEPRECATED)   \\
\hline
\multicolumn{2}{c}{\textbf{PILLARS}} \\
\hline
CWE-284 & Improper Access Control \\
CWE-707 & Improper Neutralization \\
CWE-682 & Incorrect Calculation \\
CWE-693 & Protection Mechanism Failure \\
\bottomrule
\end{tabular}
\end{table}

\subsection{\rqfivename}
\textit{RQ5: \rqfivetxt}
We found that, of the 8,033 issues analyzed, 3,697 (46\%) were resolved by the same developer that contributed the issue. The median of medians across all projects was 50\% of issues being resolved by the original contributor.

\begin{framed}
\noindent 46\% of fuzzbugs are fixed by the developer that authored the bug-contributing commit.
\end{framed}

With less than half of fuzzbugs reported by OSS-Fuzz being fixed by the author who likely caused the issue in the first place, these results indicate that many developers are not taking advantage of or being afforded a valuable learning opportunity that fuzzing feedback provides.


\section{Limitations}
We view our limitations as being internal, external, and construct. Our internal limitations are primarily concerned with how well our measurements of dates and fuzzbug counts. Our external and construct limitations are primarily about mapping fuzzbugs to their types (i.e. CWE).

\textbf{Commit range reports in OSS-Fuzz.} This refers to Step 2 of Section \ref{sec:methodology}. The OSS-Fuzz Monorail system documents all fuzzbugs found, including ones later determined to be false positives. The expected process for this is for developers to mark those as ``won't fix'', but developers do not always follow through with that. Therefore, if the code changes later on and the false positive fuzzbug is no longer observed, that commit range would be included in our analysis. We did not see this behavior as pervasive in the data set, but some commit ranges for false positive fuzzbugs do exist in our data set. Since our data set is public, future researchers can re-run the analysis as needed. 

\textbf{Identifying fix-contributing commits}. In Step 3 of Section \ref{sec:methodology}, we selected fix-contributing commits based on the mentioning of the OSS-Fuzz issue number or the presence of functions or methods logged in the Crash State as detailed in Step 1. Not every fuzzbug contained fix-contributing commits, and sometimes multiple modifications can be made within the same commit ranges. To mitigate this, we manually investigated every fuzzbug in our pilot study (i.e. \emph{systemd}) and derived our methodology in Step 3 of analyzing the crash state information from that exercise.

\begin{table*}[]
\centering
\caption{CWE Classes encountered in vulnerability type comparison, and which have been fuzzbugs }
\label{tab:cweclasses}
\begin{tabular}{l l l c}
\toprule
\textbf{CWE ID} & \textbf{CWE Class Name} & \textbf{CWE Bases, Variants IDs} & \textbf{Fuzzbug?} \\
\midrule
CWE-20 & Improper Input Validation & & No \\
CWE-74 & Improper Neut. of Special Elements in Output Used by a &  89, 91, 94, 1236, 79, 87, 80, 93, 113 & No \\ 
       & Downstream Component. ('Injection') & & No \\ 
CWE-77 & Improper Neutralization of Special Elements & 78 & No \\ 
       & used in a Command ('Command Injection') &  \\
CWE-99 & Improper Control of Resource Identifiers ('Resource Injection') & 88 & No \\
CWE-114 & Process Control & & No \\
CWE-116 & Improper Encoding or Escaping of Output & 838 & No \\
CWE-119 & Improper Restriction of Operations within the Bounds & 120, 125, 787, 788, 824 & Yes \\ 
& of a Memory Buffer & & \\
CWE-172 & Encoding Error & & No \\
CWE-200 & Exposure of Sensitive Information to an Unauthorized Actor & 201, 203, 209, 538& No \\
CWE-269 & Improper Privilege Management & 266, 268 & No \\
CWE-271 & Privilege Dropping / Lowering Errors & 273& No \\
CWE-285 & Improper Authorization & 552, 276, 281, 862, 425, 863, 639 & No \\
CWE-287 & Improper Authentication & 290, 294, 305, 306, 307, 521, 798, 522, 256 & No \\
CWE-311 & Missing Encryption of Sensitive Data & 312, 319& No \\
CWE-326 & Inadequate Encryption Strength & & No \\
CWE-327 & Use of a Broken or Risky Cryptographic Algorithm & 916 & No \\
CWE-330 & Use of Insufficiently Random Values & 331, 338 & No \\    
CWE-345 & Insufficient Verification of Data Authenticity & 346, 347, 354, 494 & No \\
CWE-362 & Concurrent Execution using Shared Resource & 364, 367 & No \\
 & with Improper Synchronization ('Race Condition') & &  \\ 
CWE-377 & Insecure Temporary File & 378, 379 & No \\
CWE-400 & Uncontrolled Resource Consumption & 770, 789 & No \\
CWE-404 & Improper Resource Shutdown or Release & 459, 763, 772, 401 & No \\
CWE-436 & Interpretation Conflict & 444 & No \\
CWE-451 & User Interface (UI) Misrepresentation of Critical Information & 1021& No \\
CWE-522 & Insufficiently Protected Credentials & 256& No \\
CWE-610 & Externally Controlled Reference to a Resource in Another Sphere & 73, 601, 611, 384 & No \\
CWE-665 & Improper Initialization & 789, 908, 1187, 909, 1188 & Yes \\
CWE-667 & Improper Locking & & No \\
CWE-668 & Exposure of Resource to Wrong Sphere & 22, 134, 427, 428, 552,& No \\
CWE-669 & Incorrect Resource Transfer Between Spheres & 212, 434, 494, 829& No \\
CWE-670 & Always-Incorrect Control Flow Implementation & 617 & Yes \\
CWE-672 & Operation on a Resource after Expiration or Release & 613& No \\
CWE-674 & Uncontrolled Recursion & & No \\
CWE-704 & Incorrect Type Conversion or Cast & 681, 843, & Yes \\
CWE-706 & Use of Incorrectly-Resolved Name or Reference & 22, 59, 64, 178& No \\
CWE-732 & Incorrect Permission Assignment for Critical Resource & 276, 281 & No \\
CWE-754 &  Improper Check for Unusual or Exceptional Conditions & 252, 273, 354, 476 & Yes \\
CWE-755 & Improper Handling of Exceptional Conditions & 209 & No \\
CWE-834 & Excessive Iteration & 835 & No \\
CWE-913 & Improper Control of Dynamically-Managed Code Resources & 94, 502, 915, 1321 & No \\
CWE-922 & Insecure Storage of Sensitive Information & 312 & No \\
        & No Class & 369, 121, 1901, 415, 416, 129, 190, 248 & Yes \\
        & No Class & 23, 36, 61, 122, 123,  131, 184, 191, 193,  & No \\ 
        &          & 358,  426, 471, 532, 548, 565, 749, 769, 918, 1333 &  \\

\bottomrule
\end{tabular}
\end{table*}
\textbf{Identifying origin commits.} In Step 4 of Section \ref{sec:methodology}, identifying the origin commit from a fix commit is an imperfect process. We used tooling based on a long history of refining and expanding upon the SZZ algorithm \cite{szz2005}, which has been widely used in the mining software repository \cite{neto2018, szz2006, alohaly2017} research community and its benefits and drawbacks are well-understood. We used the tool archeogit deveoped by Munaiah et al. \cite{munaiah2019} to perform a \verb|git blame| operation with a provided fix commit as a base.  Further, if the fix-contributing commit implemented a workaround to hide the issue, the blame operation likely selected an incorrect bug-contributing commit. 

\textbf{Mapping to CWE}. One construct limitation is in the CWE mapping in our data analysis in Section \ref{Analysis}. The mapping of Crash Types as detailed in Section \ref{analysis-4} to CWEs was a subjective task. To mitigate this limitation, we erred on the side of broader mappings and are fully transparent with our choices. By providing our mapping in this paper and the availability of our data, anyone can make corrections to our mapping and re-compute our results.

\textbf{CWE generalizability}. One external limitation is that the CWE itself is not comprehensive, and is subject to expand over time as it is an ongoing project. The CWEs utilized in the mapping were strictly those that the original reporters to the CVE chose to select, since the CWE identifier is optional in the NVD. This could lead to underestimation of the overlap between CVEs and OSS-Fuzz fuzzbugs. We used CWE version 4.8, the latest version of CWE available at the time of this writing.

\section{Summary}
The goal of this work is to empower developers in determining if fuzzing is a worthwhile investment in their development process. We investigated five research questions spanning lifespans, response times, frequency, relevance, and learning opportunities. We analyzed fuzzbugs as filed with Monorail and the commits contributing to their introduction and rectification. We further used the CVE records for the projects that use OSS-Fuzz to explore the overlap in bug types. Our results indicate that fuzzers help to reduce the time between bug introduction and discovery. Further, our results show that developers respond quickly to fuzzbugs and often are given the opportunity to correct their own mistakes. Our results also show that fuzzbug overlap with CVEs is not extensive, suggesting that, while OSS-Fuzz helps to locate a wide variety bugs, they are far from comprehensive in terms of a security audit. Overall, these results provide detailed assistance to software development teams to decide whether adopting fuzz testing as a practice makes sense for their project.

\section*{Acknowledgement}
This work has been supported by the National Science Foundation Cybercorps Scholarship for Service (grant 1922169) and the Department of Defense DARPA program. This work also was supported by the Vulnerability History Project. Any opinions, findings, and conclusions or recommendations expressed in this material are those of the author(s) and do not necessarily reflect the views of the National Science Foundation or the Department of Defense DARPA program.

\bibliographystyle{ieeetr}
\bibliography{references}

\end{document}